# Study of Timing Synchronization in MIMO-OFDM Systems Using DVB-T

Farhan Farhan

Electrical Systems Technology Instructor, A-B Tech, Asheville NC, USA

*ABSTRACT*

*OFDM (Orthogonal Frequency Division Multiplexing)provides the promising physical layer for 4G and 3GPP LTE Systems in terms of efficient use of bandwidth and data rates. This paper highlights the implementation of OFDM in Digital Video Broadcasting-Terrestrial (DVB-T). It mainly focuses on the timing offset problem present in OFDM systems and its proposed solution using Cyclic Prefix (CP) as a modified SC (Schmidl and COX) algorithm. It also highlights the timing synchronization as well as performance comparison through bit error rate. Synchronization issues in OFDM are important and can lead to information loss if not properly addressed. Simulations were performed to implement DVB-T system and to compare different synchronization methods under certain distribution model.*

## 1. INTRODUCTION

The advancement in wireless and mobile communication has led the researchers to design and implement transceiver with higher data rates using the bandwidth efficiently. The single carrier modulation scheme is not enough to meet the demands of high data rates and higher capacity. To achieve this goal, multicarrier networks such as FDM (Frequency Division Multiplexing) were utilized since 1950's but it was hard to implement those networks. This implementational complexity caused to limit their use for military applications[1]. In multicarrier system, there are a number of carriers which carry the information in parallel manner. This system is less susceptible to noise and interference especially in wireless environment. Chang [2] was the first who introduced Orthogonal Frequency Division Multiplexing (OFDM) by further developing and utilizing the Frequency Division Multiplexing (FDM) in 1950. In OFDM systems, synchronization at the receiver is one of the important steps to retrieve the information correctly. There are two ways to achieve the synchronization: one is to find the correct incoming symbol timing and the other is to determine the carrier frequency offset (CFO). This paper discusses the synchronization approach to correct timing offset with the assumption that CFO correction has been made. There are many approaches found in literature to observe the timing synchronization in conventional OFDM systems. This paper investigates the effect of timing offset on Bit Error Rate Probability to determine the performance analysis of OFD Mutilized in DVB-T scheme. The European Telecommunication Standards Institute (ETSI)adopted a standard called DVB-T in 1997 for digital information transmission through terrestrial channels. OFDM was the chosen modulation approach in this standard and there are two modes for the implementation of DVB-T namely, 2K and 8K mode.

## 2. OFDM SYSTEM MODEL

A basic OFDM system model [3] has PSK or QAM modulator/demodulator, a serial to parallel /parallel to serial converter, and an IFFT/FFT module. The hardware implementation of OFDM requires the use of either ASIC (Application Specific Integrated Circuit) or FPGA (Field Programmable Gate Array) in applications such as DVB, DAB and SDR (Software Defined Radio) and so on. Mathematically, an $l_{th}$ OFDM signal at the $k_{th}$ subcarrier can be expressed as [4]

$$\psi_{l,k}(t) = \begin{cases} e^{j2\pi f_k(t-lT_s)}, & 0 < t \le T_{sym} \\ 0, & elsewhere \end{cases}$$

Then the continuous-time passband OFDM signal is given by

$$x_l(t) = Re\left\{ \frac{1}{T_{sym}} \sum_{l=0}^{\infty} \left\{ \sum_{k=0}^{N-1} X_l[k]\psi_{l,k}(t) \right\} \right\}$$

Therefore, the discrete-time baseband OFDM signal can be achieved by sampling the continuous-time OFDM signal at $t = lT_{sym} + nT_s$ and is given as

$$x(n) = \frac{1}{\sqrt{N}} \sum_{k=0}^{N-1} X_k e^{j2\pi kn/N} \qquad -N_g \le k \le N + N_g$$

Where N is the IFFT/FFT length. $N_g$ represents the number of guard samples.
Considering perfect timing and no carrier frequency offset, the samples at receiver can be expressed as

$$r(k) = x(k) e^{\frac{j2\pi k}{N}} + n(k)$$

where $n(k)$ is the additional Gaussian white noise.

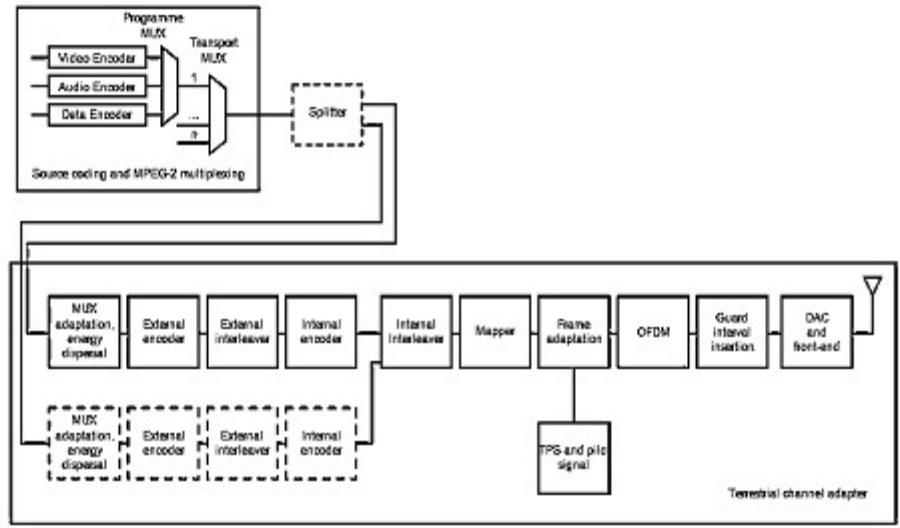

Figure 1 : Block Diagram for DVB-T Transmitter

## 2.1. OFDM Transmission and Reception in DVB-T

The block diagram for DVB-T standard is shown in the Figure 1. The processes described in this diagram are carried out in a Digital Signal Processor (DSP). This paper focuses on the implementation of 2K mode of DVB-T standard [5] and the parameters for this mode are given in table.

| Parameters | 2K Mode |
| --- | --- |
| Elementary Period (T) | 7/64 $\mu s$ |
| Number of Carriers (K) | 1705 |
| Value of carrier number $K_{min}$ | 0 |
| Value of carrier number $K_{max}$ | 1704 |
| Duration ($T_U$) | 224 $\mu s$ |
| Carrier Spacing (1/$T_U$) | 4464 Hz |
| Spacing between carriers $K_{min}$ and $K_{max}$ | 7.61 MHz |
| Allowed Guard Interval $\Delta/T_U$ | 1/4, 1/8, 1/16, 1/32 |
| Duration of Symbol part $T_U$ | 224 $\mu s$ |
| Duration of guard interval $\Delta$ | 56 $\mu s$, 28 $\mu s$, 14 $\mu s$, 7 $\mu s$ |
| Symbol Duration $T_s = \Delta + T_U$ | 280 $\mu s$, 252 $\mu s$, 238 $\mu s$, 231 $\mu s$ |

Table 1 : OFDM Parameters for DVB-T 2K Mode

The OFDM signal expression in DVB-T is given by the following emitted signal

$$x(t) = Re\left\{e^{j2\pi f_c t} \sum_{m=0}^{\infty} \sum_{l=0}^{67} \sum_{k=k_{min}}^{k_{max}} c_{m,l,k} \psi_{m,l,k}(t)\right\}$$

Where,
k denotes the carrier number;
l denotes the OFDM symbol number;
m denotes the transmission frame number;
K is the number of transmitted carriers;
Ts is the symbol duration;
$T_U$ is the inverse of the carrier spacing;
∆ is the duration of the guard interval;
$f_c$ is the central frequency of the radio frequency (RF) signal;
k is the carrier index relative to the center frequency,
$k' = k − (Kmax + Kmin)/2$
$c_{m, 0, k}$ complex symbol for carrier k of the Data symbol no.1
in frame number m;
$c_{m, 1, k}$ complex symbol for carrier k of the Data symbol no.2
in frame number m;
...
$c_{m, 67, k}$ complex symbol for carrier k of the Data symbol
no.68 in frame number m.

The 2K mode of DVB-T is specifically for the mobile reception of standard definition DTV (Digital Television).The OFDM transmission is carried out in frames. Each frame has duration of $T_F$ and consists of 68 OFDM symbols.One super-frame consists of four frames and each symbol is comprised of 1705 carriers in 2K mode with Ts as the symbol duration. It is composed of two parts: a useful part with duration $T_U$ and a guard interval with a duration ∆.

The guard interval consists in a cyclic continuation of the useful part, $T_U$, and is inserted before it. There are standards only for DVB-T transmitter and the design of receiver is open.

## 3. TIMING SYNCHRONIZATION AND PROPOSED APPROACH

In OFDM, modulation and demodulation are done using IFFT (Inverse Fast Fourier Transform) and FFT (Fast Fourier Transform) techniques, which also cause delay in the system. It is important to find the start of the OFDM symbol at the receiving end and the error in estimating the beginning of symbol causes the timing offset error. Therefore it is necessary to perform symbol time synchronization to find the start of the OFDM symbol. If this delay is not given any attention, it causes phase rotation in the constellation. If δ is the amount of offset in time domain then it causes a phase offset of 2πkδ/N in the frequency domain. This phase offset varies directly in proportion with the sub-carrier index k. There are several methods to address the issue of timing offset correction in literature. Classen focuses a method to find both timing offset and carrier frequency offset together. In his method, he considered the trial and error method, which is very computationally complex [6]. Another way to find the timing offset is using the means of slopes of the curve of carrier phases in pilots versus the index of pilots. Minn and Bhargava presented the timing synchronization by modifying the Schmidl and Cox's method and using a new pattern for training sequence [7].This paper provides the modified Schmidl and Cox's approach by not using the training sequence and making cyclic prefix (CP) as the reference. The main advantage of avoiding training sequence is to obtain efficiency in transmitting power. The methods which require the use of training preamble requires extra overhead for transmitting training sequences along with the information signal. This approach uses the sliding window technique to compute the correlation of the received signal with the cyclic prefix, which is already known,to the user. The timing metric obtained by the correlation shows the variance and the start of OFDM frame can be taken where the maximum lies in this spread of variance.

## 4. SIMULATIONS RESULTS

The DVB-T 2K mode was simulated considering the effect of Additive White Gaussian Channel. The simulations were carried out to observe the effect of timing offset along with different channel environment. These results also include the timing metric and correlation of the samples to correct timing offset error. The Bit error probability (BER) plots were also observed to see the trade off between SNR and BER with effect of timing offset.It is clear that carriers are baseband discrete-time and we can use them for discrete-time baseband simulation. However, the focus of this implementation was on passband continuous time-domain simulation. The first step to produce a continuous time-domain signal is to apply a transmit filter g(t).After using the transmit filter, a reconstruction filter D/A was used to avoid aliasing. The proposed reconstruction filter[5] is a Butterworth filter of order 13 and cut-off frequency of approximately 1/T.

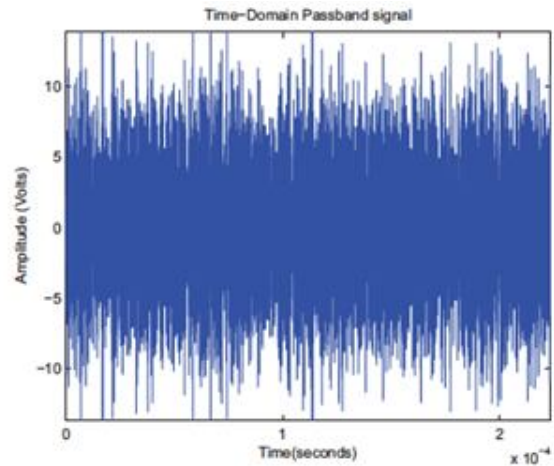

Figure 2: Passband Signal

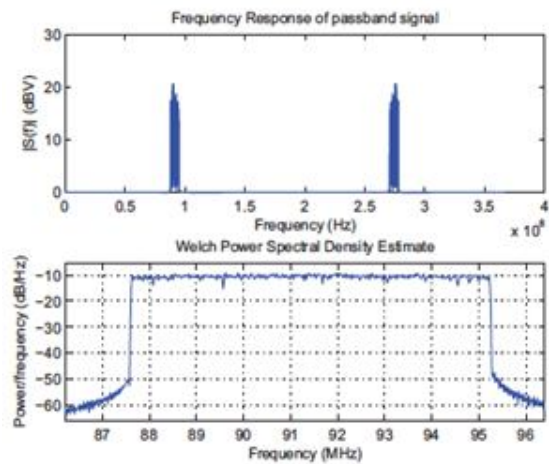

Figure 3: Frequency Response of the Passband Signal

Figure 3 shows the frequency response of the time-domain passband signal as well as power density spectrum of the signal shown over the useful bandwidth.

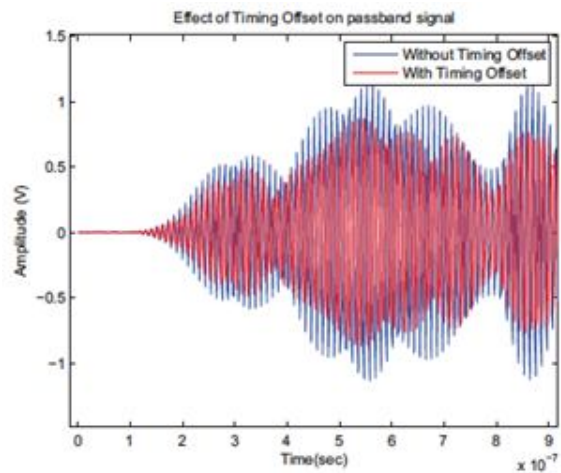

Figure 4: Time Response of the Signal After Timing Jitter

The effect of timing offset in passband signal can be seen in Figure 4 where the comparison between signal with and without time offset is shown.

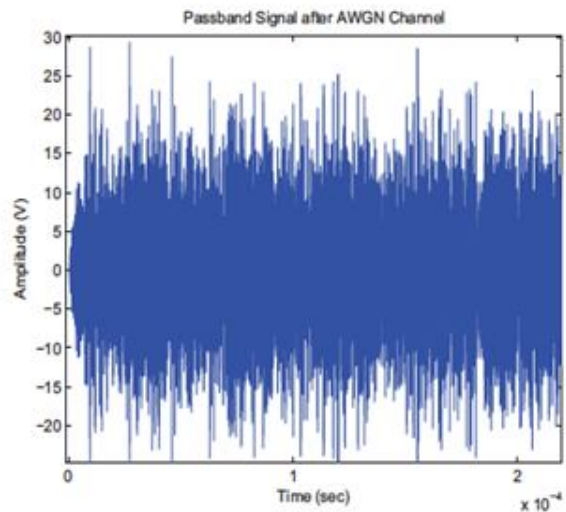

Figure 5: Effect of AWGN Channel

To observe the effect of timing offset in the system, physical channel information was added. The distribution of channel used was AWGN. The effect of AWGN channel can be seen in Figure 5.

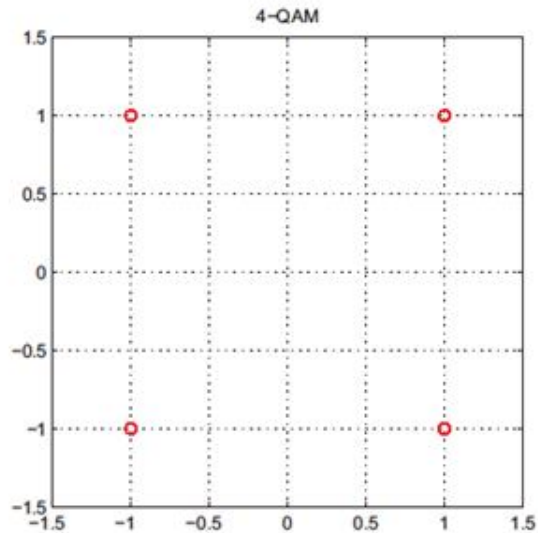

Figure 6: Signal Carriers at Receiver

The implementation of receiver was reverse of transmitter. The plot shown in Figure 6 shows stem plot of the carriers received at the receiver side.

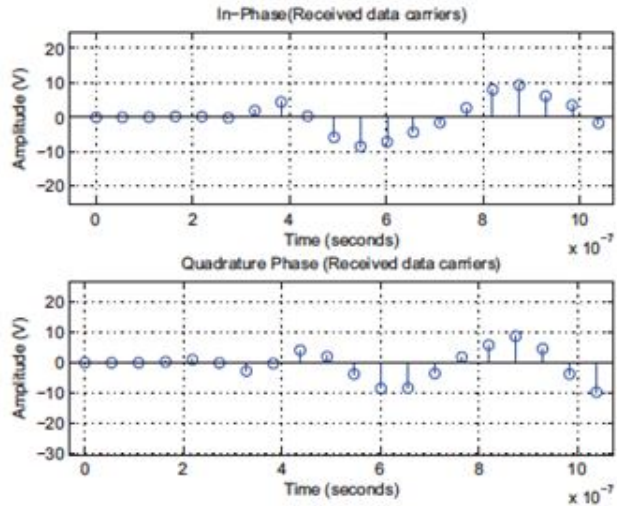

Figure 7: Ideal Constellation Diagram

The ideal constellation diagram of a 4-QAM signal is also shown in Figure 7. The study was carried out by considering one distribution models namely, AWGN. In this study, the sampling rate is defined by Ts =2.747ns and the time offset was measured by the number of samples delayed. Hence, the actual delay can be calculated using Td = kTs where k is the number of samples delayed.

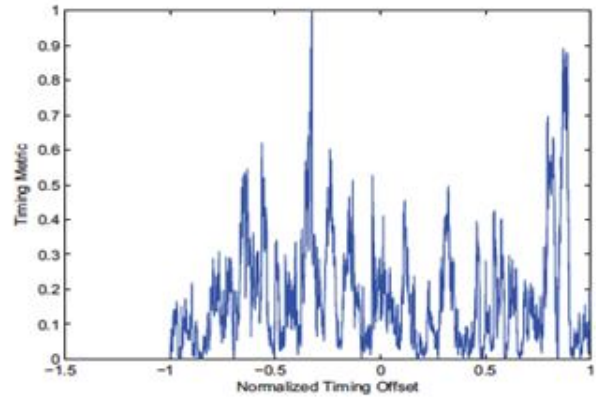

Figure 8: Proposed Approach

After implementing the proposed approach, the result of timing metric (Figure 8) is obtained in terms of correlation of samples. The effect of noise imposed by the channel was assumed to be cancelled at the receiver side. The case-by-case analysis was performed where each case was studied with different value of time-offset samples. It is clear that the reconstructed signal constellation with higher SNR value follows the ideal constellation more closely than those with less SNR value.

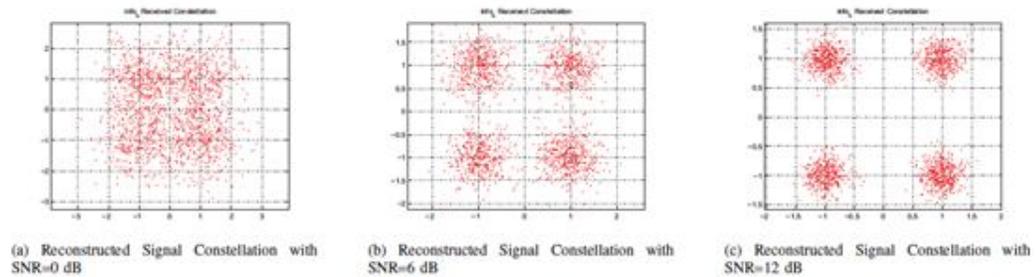

Figure 9: Case-I (Time Offset=1 Sample)

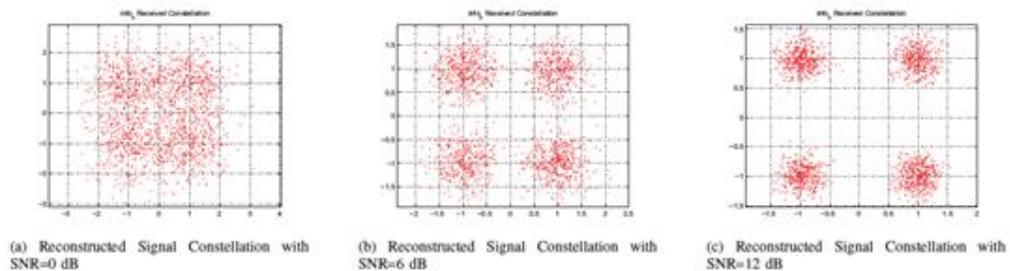

Figure 10: Case-II (Time Offset=5 Samples)

Figure 9 and 10, shows the comparison of different cases where the value of time offset sample varies. The BER Plots (Figure 11) show that SNR value required to correct the bit error rate is higher in case of 5 samples delay than with 1 sample delay.

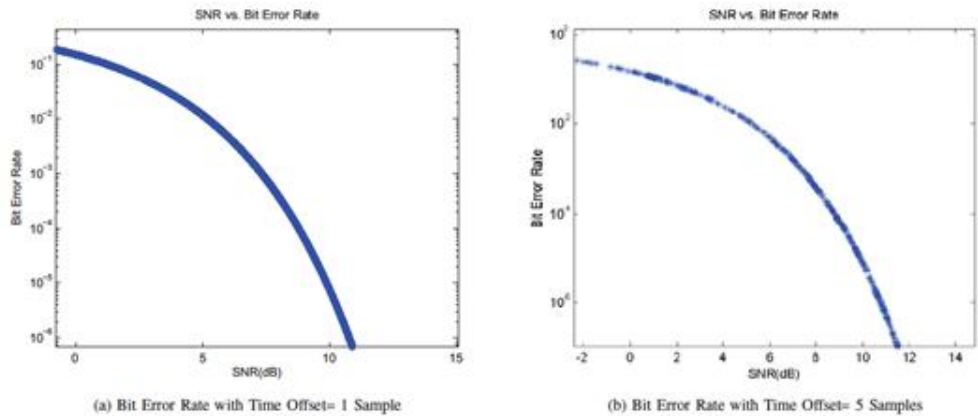

Figure 11: BER Comparison

Simulating the model with different conditions in AWGN environment performed the comparison. This comparison is shown in Table 2 in which the trend of increasing time offset on SNR in dB was observed.

| Timing Offset | BER | SNR (dB) |
| --- | --- | --- |
| 1 | $10^{-5}$ | 9.8 |
| 5 | $10^{-5}$ | 10 |
| 15 | $10^{-5}$ | 10.5 |

Table 2: Comparison of Time Offset with SNR in dB

## 4. CONCLUSIONS

The main objective of this paper was to study and investigate the need for timing synchronization errors in OFDM system, the major effects of timing offset errors, performance of system with reference to BER and then to compare some of the synchronization algorithms. The goal was achieved by implementing the OFDM system using DVB-T as an example with the effect as well as correction of timing jitter. On the concluding remarks, it was observed that AWGN environment was primitive and good enough to study the topic whereas Fading environment gives the good picture of practical scenario. The future work will involve the investigation and simulation of different probabilistic models in state of the art technologies such as 5G LTE.

## Authors

Mr. Farhan Farhan has been in the research field of communication and wireless technology for more than four years. He has received his Master of Science degree in Technology (Electrical Engineering) from Western Carolina University, Cullowhee, North Carolina, USA in 2013 and Bachelor of Science degree in Electrical Engineering from University of Engineering and Technology, Lahore Pakistan.

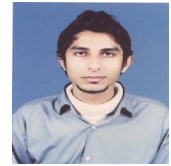